\documentclass[preprint,12pt]{elsarticle}

\usepackage{graphicx}
\usepackage{amssymb}

\usepackage{lineno}
\usepackage{enumitem}
\usepackage{algorithm}
\usepackage{algpseudocode}
\usepackage{amsmath}
\usepackage{hyperref}

\newcommand{\proref}[1]{Protocol~\ref{#1}}

\newcommand{\funcref}[1]{Functionality~\ref{#1}}

\newcommand{\A}{{\ensuremath{\mathbb A} }} 
\newcommand{\F}{{\ensuremath{\mathbb F_p} }}

\makeatletter
\newsavebox{\mybox}
\newcounter{pro}
\newcounter{func}

\newenvironment{pro}[1]{
\renewcommand{\ALG@name}{Protocol}
\let\c@algocf\c@pro
\begin{algorithm}[#1]%
}%
{
\end{algorithm}}

\makeatletter

\newenvironment{func}[1]{
\renewcommand{\ALG@name}{Functionality}
\setcounter{func}{\value{algorithm}} 
\setcounter{algorithm}{\value{pro}}%
\begin{algorithm}[#1]%
}%
{
\end{algorithm}
\setcounter{pro}{\value{algorithm}}
\setcounter{algorithm}{\value{func}}
}

\makeatother
 
\usepackage{xargs}                      
\usepackage{amsthm}

\usepackage{float}
\usepackage[colorinlistoftodos,prependcaption,textsize=small]{todonotes}
\newcommandx{\rt}[2][1=]{\todo[linecolor=red,backgroundcolor=red!25,bordercolor=red,#1]{#2}}

\journal{arXiv}

\begin{document}

\begin{frontmatter}


\title{ Privacy Preservation in Epidemic Data Collection }




\author[1]{Katrine Tjell}
\author[2]{Jaron Skovsted Gundersen}
\author[1]{Rafael Wisniewski}

\address[1]{Department of Electronic Systems, Aalborg University, Denmark}
\address[2]{Department of Mathematical Sciences, Aalborg University, Denmark}
\begin{abstract}
This work is inspired by the outbreak of COVID-19, and some of the challenges we have observed with gathering data about the disease. To this end, we aim to help collect data about citizens and the disease without risking the privacy of individuals. Specifically, we focus on how to determine the density of the population across the country, how to trace contact between citizens, how to determine the location of infections, and how to determine the timeline of the spread of the disease. Our proposed methods are privacy-preserving and rely on an app to be voluntarily installed on citizens' smart-phones. Thus, any individual can choose not to participate. However, the accurateness of the methods relies on the participation of a large percentage of the population. 
\end{abstract}


\end{frontmatter}


\section{Introduction}
When an epidemic breaks out, breaking the chain of infection is one of the vital defends against the spread of the disease. Besides that, it is essential to know how the population is distributed across the country at different times, and where and when most people become infected. This kind of information can be crucial for the authorities to plan how the society can get through the epidemic.

In this paper, we strive to answer such questions by using the population itself. The main idea is based on the fact that a large percentage of citizens carry a smart-phone around with them. This smart-phone can be used to collect and share data that can be helpful in the defeat of epidemic diseases. At the same time, we aim to preserve the privacy of each citizen, such that individuals will not be forced to reveal private information. We will present technical solutions based on cryptographic methods.

The main scenario is a country where most of the citizens carry a smart-phone at all times. When we write that each citizen performs an action, we mean that an app on her smart-phone does this action automatically. Moreover, we assume there is one or two servers available that do not collude. One could imagine that a tele-service provider/operator owns one server, and the health-authoritie owns the other server. To this end, we assume that all smart-phones can privately communicate with both servers. This could be achieved using an encrypted internet connection. We assume that the servers follow the instructions, but we do not trust them with private information. We also assume that the users of the app (the citizens) follow instructions, however, we do discuss how a user might cheat and suggest ways to prevent this. In this scenario, we will consider the following challenges:
\begin{enumerate}[label={(\bfseries P\arabic*)}]

    \item Determine the distribution of citizens in geographic locations.
    \item Trace contacts with infected citizens. That is, how to accelerate the process of identifying (potentially) infected people?
    
    \item Determine geographic locations of infections.
    
    \item Determine the time it takes for the disease to spread from one geographic location to another.  
    

\end{enumerate}


    
    


\subsection{Related work}
Since the outbreak of COVID-19, several research groups have been occupied with the development of technology that can aid in containing an epidemic disease. The newly founded PEPPPT organization, \cite{PEPPPT}, is developing solutions to stop the spread of COVID-19 while focusing on the privacy and data protection laws in Europe. To date, PEPPPT has proposed a privacy-preserving \textit{contact tracing} method, which is meant to aid citizens in finding out if they have been in contact with a person that has later tested positive for the disease. The method is developed for application on smart-phones. Its efficiency relies on the number of citizens that volunteer to install the app on their smart-phone. The idea is based on random numbers being transmitted and received by smart phones, such that random numbers transmitted by a person that is later tested positive for the disease can help to tell whom the infected person has been in contact with. 
Also, the work \cite{D3PT} is concerned with privacy-preserving contact tracing adhering to the privacy laws in Europe. While they propose a decentralized solution, the underlying idea is the same as in the work by PEPPPT. Another work that is based on this idea is \cite{CTV20}. 

The MIT originated work in \cite{MITapp} is also about privacy-preserving contact tracing; however, this method is not based on random numbers. The idea in this work is instead based on GPS location data, such that when a person is tested positive for the disease, redacted location data of the person can be published, and others can see if they have made contact by comparing their own location data. They claim that redacting the location data preserves the privacy of the individual.  

In our work, we consider a few more problems, \textbf{(P1)-(P4)}, in addition to contact tracing. Our method for contact tracing is built on the works \cite{D3PT, PEPPPT}, specifically, our contribution is a cryptographic technique on how to strengthen the privacy preservation of the method.

\subsection{Structure}
In sections \ref{sec:2} - \ref{sec:5}, we present our proposed methods for solving the problems \textbf{(P1)-(P4)}, respectively. Section \ref{sec:6} concludes the paper.

\section{Private Determination of the Distribution of Citizens across the Country} \label{sec:2}

The task consists of estimating the distribution of the population over a region - a set $A \subset R^2$. That is, we want to approximately determine where citizens are located in $A$ at a certain time; or more precisely, how many citizens are located at different locations of $A$. The latter underlines the fact that we do not seek to learn the location of individuals. To this end, we define a partition of $A$ as $\mathcal{A} = \{ A_j|~ j \in J = \{1,\ldots,M\}\}$, where $M$ is the number of locations, such that
\begin{equation}\label{eq:parts}
\begin{aligned} 
&\bigcup_{j \in J} A_j = A, \\
& A_i \cap A_j = \emptyset, \hbox{ for } i \neq j.     
\end{aligned}
\end{equation}
We will use the terminology \textit{location} for a set $A_j\in \mathcal{A}$.

We propose to solve the problem by calculating the number of citizens in each location $A_j$. To protect the privacy of each individual, we do not reveal the identity of the citizens in $A_j$. 

\subsection{Problem Formulation}
Let $I$ be an index set - the set of citizens carrying a smart-phone, and $x_i \in A$ be the position of citizen $i \in I$. Consider the map $P : A \to \mathbb{Z}^M$,
\begin{equation}\label{eq:P}
    P(x) = (I_{A_j}(x))_{j\in J},
\end{equation} 
where 
\begin{equation} \label{eq:I}
    I_{A_j}(x) = \begin{cases} 1 & \text{if } x \in A_j \\
0 & \text{otherwise}.
\end{cases}
\end{equation}

For a position $x$ in $A$, $P(x)$ delivers the information of which location $A_j\in \mathcal{A}$, $x$ belongs to. For instance if $x \in A_1$, $P(x) = (1,0,0\cdots,0)$. The problem is then to calculate
\begin{equation} \label{eq:sum}
    \sum_{i \in I} P(x_i),
\end{equation}
which yields the number of citizens in each location of the area. We aim to calculate this without revealing the location of individuals.

\subsection{Protocol} The main idea behind the protocol is for each citizen $i \in I$ to send their encrypted information $[P(x_i)]$ to the server, using their smart-phone. The server should then compute
\begin{align} \label{eq:encSum}
   [D] = \sum_{i \in I} [P(x_i)],
\end{align}
which can be done directly on the encrypted data, using for instance
the Paillier encryption scheme. There are a few challenges with the proposed idea; 1) if $M$ is large, i.e., $A$ is divided into many locations, then $P(x_i)$ will be a large vector. Thus, sending $[P(x_i)]$ might be impractical given the size of $M$. 2) We want to be able to decrypt $D$ in \eqref{eq:encSum}, while at the same time ensuring that each individual term $[P(x_i)] $ cannot be decrypted.
Considering the first of these challenges, we propose a trade-off between privacy and complexity by letting each citizen $i \in I$ send only a part of $P(x_i)$ together with the index information. That is, citizen $i$ chooses $\bar{M} < M$ indices $j_i \in J$ where the index of the location she is located in is one of them, and the indices form the vector 

\begin{equation}\label{eq:vi}
    J_i = (j_1, \ldots, j_{\bar{M}}),
\end{equation} 
For a vector $v \in \mathbb{N}^{\bar{M}}$ and the partition $\mathcal{A}$, we define the map $I_{A(v)} = (I_{A_{v_1}}, \hdots, I_{A_{v_{\bar{M}}}})$.
    
Let $P_i: A \to \mathbb{Z}^{\bar{M}}$ be the map defined by 
$$
P_i(x) = I_{A(J_i)}(x).
$$
The citizen $i$ then sends $(J_i, [P_i(x_i)])$ to the server.
With this approach, citizen $i$ reveals that she is in one of the $\bar{M}$ locations in $J_i$, thus a large $\bar{M}$ gives better privacy but also more complexity.   

To address the second challenge, we consider the scenario where there are two non-colluding servers available. In this scenario, the protocol for privately computing \eqref{eq:sum} is simple, but the privacy of the protocol relies on trusting that the two servers do not collude. We present this protocol in \proref{pro1}.

\begin{pro}{htbp}
	\caption{Determination of Distribution of Citizens}
	\label{pro1}
		\begin{algorithmic}[1]
		    \Statex $p > |I|$ is a prime and $\F$ is a finite field with $p$ elements. 
		    \vspace{0.2cm}
		    \Statex \textbf{\textit{Each citizen $i \in I$ does:}}
		    \State Choose $J_i$ according to \eqref{eq:vi}.
            \State Choose $r_i \in \F^{\bar{M}}$ uniformly at random.
            \State Compute $y_i = (P_i(x_i) + r_i) \mod p$.
            \State Send $(J_i, r_i)$ to server 1 and $(J_i, y_i)$ to server 2.
            
            \Statex \textbf{\textit{Server 1 does:}}
            
            \State Compute for all $i \in I$: \begin{align}
                r'_i[J_i[k] ] &= r_i[J_i[k]] && \text{for } k=1,\ldots, \bar{M},\\
                r'_i[k] &= 0 && \text{for } k \neq J_i[h] \: \forall\: h.
            \end{align}
            where $x[k]$ means the $k$'th entry in the vector $x$.

            \State Compute $$s_1 = \left( \sum_{i\in I} r'_i \right) \mod p.$$
            \State Send $s_1$ to server 2.
            
            \Statex \textbf{\textit{Server 2 does:}}
            \State Compute \begin{align}
                y'_i[J_i[k]] &= y_i[J_i[k]]  &&\text{for } k=1,\ldots, \bar{M}, \\
                y'_i[k] &= 0 && \text{for } k \neq J_i[h] \: \forall \: h.
            \end{align}
            \State Compute $$D = \left( \sum_{i\in I} y'_i - s_1\right) \mod p.$$
		\end{algorithmic}
\end{pro}

The correctness of \proref{pro1} follows from the following;
\begin{equation}
    \begin{aligned}
        D &= \left( \sum_{i\in I} y'_i - s_1 \right)\mod p \\
        &=  \left( \sum_{i\in I} P(x_i) + r_i'\right) \mod p - \left( \sum_{i\in I} r_i' \right) \mod p \\
        &= \sum_{i\in I} P(x_i) + \left(\sum_{i\in I} r_i' \right) \mod p - \left( \sum_{i\in I} r_i' \right) \mod p \\
        &= \sum_{i\in I} P(x_i),
    \end{aligned}
\end{equation}
which shows that server 2 leans the sum in \eqref{eq:sum} as required. To see that the protocol is privacy preserving with respect to citizen $i$, consider the values server 1 and server 2 receives during execution. Server 1 receives from citizen $i$ the data $(J_i, r_i)$, where $J_i$ is a vector of $\bar{M}$ locations and $i$ is in one of them. $r_i \in \F^{\bar{M}}$ is a vector of uniformly random numbers giving no information to server 1. Thus, from $J_i$, server 1 can with probability $\frac{1}{\bar{M}}$ guess the location of $i$.
Server 2 receives the data $(J_i, y_i)$ from citizen $i$ and $s_1$ from server 1. Both $y_i \in \F^{\bar{M}}$ and $s_1\in \F^{\bar{M}}$ are vectors of uniformly random numbers, thus these give no information to server 1. $J_i$ gives server 2 the same information that it gave server 1. 
This means that the exact location citizen $i$ is located in, cannot be inferred from the protocol.

We remark that we expect that the parties follows the instructions in the protocol since a corrupt $i$ could easily mess up the computation by choosing $P_i(x_i)$ different from a standard unit vector. This means that this protocol considers passive corruptions. However, we also notice that no information is sent to the citizens meaning that they cannot gather additional information about other citizens by combining their information.

Using \proref{pro1}, where the tele-operator takes the role of server 1 and the health-authorities take the role of server 2, results in the latter to learn the sum in \eqref{eq:sum}, without learning the location of individuals. For learning $D$ at times $t_1,t_2,\ldots$, \proref{pro1} is simply executed at each of these times. To this end, we remark that each citizen $i \in I$, should change $J_i$ in \eqref{eq:vi} as little as possible between executions, since the overlap between $J_i$ at time $t_1$ and at time $t_2$ may narrow down the possible locations of $i$. 

\section{Identification of contacts} \label{sec:3}
The task is to determine if a citizen had contact with an infected citizen. The aim of it is to warn about the possibility of being infected and advise for self-quarantine or inform that the citizen should get tested.

\subsection{Problem Formulation and Setup}
We register information of a citizen $i$ being in the vicinity (say 2 meters) of a citizen $j$. If at later time (within two weeks), $i$ is infected, we wish to inform each citizen $j$ that she was in the vicinity of $i$. 

Our suggested solution to the above problem is based on an app that each citizen can voluntarily download. The app will generate random tokens and update these tokens frequently. The idea is that the app will send out these tokens to all nearby phones, for instance via Bluetooth. The other citizens will store the received tokens. If at some point, a citizen is reported infected, the citizen can upload all its produced tokens to a server. Other citizens can compare its received tokens to the tokens on the server, and if there are common tokens, the citizen will learn that she has been nearby an infected person, and hence she is notified that it is recommended that she gets tested or self-quarantined. Similar app suggestions can also be found in \cite{D3PT, CTV20}.

To discuss the setup more formally, we introduce some terminology. We refer to the duration where the app broadcasts the same token as an \emph{epoch}. An epoch could for instance be one minute, meaning that the app updates the broadcasting token each minute. We will also talk about the \emph{retention time} which is the time the data needs to be stored in the phone and in the server (this could for instance be two weeks). At last, we define the \emph{update interval} to be the number of epochs between which a citizen will compare her set with the server's set (this might for instance be once per day).

We also introduce some notation. We let $I$ be defined as in section \ref{sec:2}. Each citizen $i\in I$ stores two sets on their phone; $L(i)$ contains all the produced tokens by citizen $i$ during the retention time and $R(i)$ contains all received tokens by nearby phones in the retention time. 

If $i$ is tested positive, it uploads $L(i)$ to the server. We denote the set of infected citizens by $Q$ meaning that the server stores the random tokens in the large set $L=\bigcup_{i\in Q} L(i)$. 

Once per update interval, each citizen need to test whether its set of received tokens $R(i)$ has enough in common with the set of reported infected tokens $L$. That is they need to learn some information about $R(i)\cap L$.

A clever way to produce the tokens can for instance be found in \cite{D3PT}. We will not go into details about the production of tokens in this work. For now, we simply assume that the parties produce a new random token each epoch, and thus they produce the sets $R(i)$ and $L(i)$. We furthermore notice that the tokens should be deleted from $R(i)$, $L(i)$, and $L$ after the retention time.

Instead of focusing on producing the tokens, we give new suggestions to the comparison with the server each update interval. In other words, we strive to compute information about $R(i)\cap L$ for each $i$ during each update interval in a secure way.

In other suggested protocols, the set $L$ is send to the citizen who can directly compute $R(i)\cap L$. The advantage of this could be if additional information about the tokens in $R(i)$ is stored together with the token. This could for instance be some information about the duration of contact or the strength of the signal when they were in contact which might could be used for calculating some probability of infection. However, we notice that even though the citizens' sets only include random tokens it might be possible for the citizen to identify the tokens to specific people depending on when the token is added to $R(i)$ and furthermore if additional information is stored along with the token. Hence, it might be undesirable that the citizens even learn what the intersection is, and furthermore which tokens are included in $L$ since this information might reveal to them the identity of infected citizens. So as a more privacy preserving alternative we suggest a protocol where the citizens only learn how many tokens they have in common with $L$ or maybe just is notified when the cardinality of the intersection is above some threshold. 

Our suggestion to achieve this is by using a multiparty computation protocol to compute the cardinality $|R(i)\cap L|$. 

\subsection{Sketch of Protocols using PSI-CA}
We consider the citizens $i \in I$ and a server $S$. Each citizen holds an $R(i)$ and the server holds $L$. We want that the citizen learns if $|R(i)\cap L|$ is large meaning that the citizen has a high risk of being infected. To compute $|R(i)\cap L|$ the parties can use a secure two-party computation protocol known as PSI-CA (private set intersection - cardinality). Informally speaking such a protocol considers two parties, a sender and a receiver. The receiver should learn the cardinality of the intersection and the sender should learn nothing. More formally, the functionality for PSI-CA is presented in \funcref{func:PSI-CA}.

\begin{func}{ht}
    \caption{$\mathcal{F}_{PSI-CA}$}
	\label{func:PSI-CA}
		\begin{algorithmic}[1]
		    \Statex On input a set $X$ of cardinality $n$ from the receiver and a set $Y$ of cardinality $m$ from the sender, the functionality outputs $\perp$ to the sender and $|X\cap Y|$ to the receiver.
		\end{algorithmic}
\end{func}
 
In many of the implementations of this functionality $|X|$ and $|Y|$ is taken as inputs meaning that the set sizes are also revealed (or at least an upper bound of the sizes). 
 
Several protocols implementing the functionality in \funcref{func:PSI-CA} can be found in the literature, see for instance \cite{DGT12, DD15, DD15b, LYYFLLLZL119}. In \proref{pro:PSI-CA}, we give a simplified version of the protocol presented in \cite{LYYFLLLZL119} which uses what is called a commutative encryption scheme. A commutative encryption scheme is an encryption scheme where the encryption function $Enc\colon M \times K \to M$ satisfies
\begin{align*}
    Enc_{k}(Enc_{k'}(m))=Enc_{k'}(Enc_{k}(m))    
\end{align*}
for all encryption keys $k,k'\in K$. Such encryption schemes can for instance be found in \cite{SRA81, PH78}. We furthermore notice that \cite{DGT12, DD15, LYYFLLLZL119} offer security against semi-honest adversaries (even though some of them offers active security against either the sender or the receiver) while \cite{DD15b} offers security against malicious adversaries. 

We only want to give an idea of how to securely compute the cardinality of the intersection. \proref{pro:PSI-CA} is presented in order for the reader to get familiar with the ideas used in the protocols from \cite{DGT12, DD15, DD15b, LYYFLLLZL119}. 

We also remark that \cite{LYYFLLLZL119} uses a Bloom filter to decrease the communication complexity but for simplicity we have left this part out. 
\begin{pro}{ht}
	\caption{PSI-CA protocol}
	\label{pro:PSI-CA}
		\begin{algorithmic}[1]
		    \Statex The receiver and the sender setup the encryption and decryption keys. We use the notation $Enc_r$, $Dec_r$, $Enc_s$, and $Dec_s$ to denote the encoding and decoding under the encoding/decoding keys of the receiver and sender respectively (we remark that the encoding and decoding keys needs not to be the same). The receiver holds $X=\{x_1,x_2,\ldots,x_n\}$ and the sender holds $Y=\{y_1,y_2,\ldots,y_m\}$. 
		    \vspace{0.2cm}
		    \State The receiver sends 
		    \begin{align*}
		        \{Enc_r(x_i)\mid i=1,2,\ldots,n\}
		    \end{align*} 
		    to the sender.
            \State The sender computes $Enc_s(Enc_r(x_i))$
            use a random permutation $\pi\colon \{1,2,\ldots,n\} \to \{1,2,\ldots,n\}$ to obtain the set 
            \begin{align}\label{eq:double_encoding}
                \{Enc_s(Enc_r(x_{\pi(i)})) \mid i=1,2,\ldots,n\}
            \end{align}
            and sends this set to the receiver along with the set
            \begin{align*}
                Y'=\{Enc_s(y_j)\mid j=1,2,\ldots, m\}
            \end{align*}
            \State The receiver can use $Dec_r$ on each element in the set from \eqref{eq:double_encoding} to obtain the set
            \begin{align*}
                X'=\{Enc_s(x_{\pi(i)})\mid i=1,2,\ldots,n\}.
            \end{align*}
           \State The receiver outputs $|{X}'\cap {Y}'|$
		\end{algorithmic}
\end{pro}

To check the correctness of this protocol, denote a permutation on $n$ elements by $\pi$, and notice that
\begin{align*}
    X'= \{Enc_S(x_{\pi(i)})\mid i=1,2,\ldots, n\}             =\{Enc_S(x_{i})\mid i=1,2,\ldots, n\},
\end{align*}
Because $Enc_s$ needs to be an injective function, in order to be able to decrypt, it follows that $|X\cap Y|=|X'\cap Y'|$. The privacy of the protocol follows by the encryption scheme. 

We now give two suggestions of how the citizens can use the $\mathcal{F}_{PSI-CA}$ functionality to check if they have been in vicinity of too many infected persons. 

\textbf{Suggestion 1:} The citizen takes the role of the receiver and the server takes the role of the sender in Protocol \ref{pro:PSI-CA}. The citizen learns $|R(i)\cap L|$ and if the cardinality of the intersection is large enough the citizen knows that she needs to get tested. This description is presented in \proref{pro:Notify1}.

\begin{pro}{ht}
	\caption{Identification of Contacts -- Suggestion 1}
	\label{pro:Notify1}
		\begin{algorithmic}[1]
		    \State The citizen and the server calls the functionality $\mathcal{F}_{PSI-CA}$ where the citizen takes the role as the receiver inputting $X=R(i)$. The server takes the role as the sender and inputs $Y=L$. The functionality outputs $|R(i)\cap L|$ to the citizen.
		\end{algorithmic}
\end{pro}

\textbf{Suggestion 2:} The citizen takes the role of the sender and the server takes the role of the receiver in Protocol \ref{pro:PSI-CA}. The server learns $|R(i)\cap L|$ and if $|R(i)\cap L|>t$ for some predetermined threshold $t$ the server sends out a notification to the citizen. This description is presented in \proref{pro:Notify2}.

\begin{pro}{ht}
	\caption{Identification of Contacts -- Suggestion 2}
	\label{pro:Notify2}
		\begin{algorithmic}[1]
		    \State The citizen and the server calls the functionality $\mathcal{F}_{PSI-CA}$ where the server takes the role as the receiver inputting $X=L$. The citizen takes the role as the sender and inputs $Y=R(i)$. The functionality outputs $|R(i)\cap L|$ to the server.
		    \State If $|R(i)\cap L|>t$ the server sends $1$ to the citizen. Otherwise it sends $0$ to the citizen.
		\end{algorithmic}
\end{pro}

\subsection{Security Advantages of using PSI-CA for Comparison}
We observe that it is unavoidable that a corrupt citizen will store additional information when receiving tokens in the discussed app. In that way, the citizen might be able to identify the received random token with some specific person if she only received one token during an epoch. If the set held by the server is public or sent to citizens it will be enough for the citizen to only know which token corresponds to which person in order to learn if this person is infected by simply comparing to the server's set.

We avoid this in some sense by introducing the functionality $\mathcal{F}_{PSI-CA}$ to carry out this comparison. We note, however, that if we are using Suggestion 1 from above it is still an opportunity that a corrupt party can learn this information using the following approach. She let $R(i)$ contain only the single token which she knows corresponds to the specific person. If she learns that the intersection is $1$ she has learned that the person is infected. This might be fixed by giving restrictions on when a comparison takes place such as, $|X|>s$ for some $s$ before the comparison takes place (we remark that this seems easy to implement if we are willing to reveal $|R(i)|$ to the server which some of the implementations already do), or alternatively use a third party to authorize the citizens set before comparison. There are also protocols for authorized private set intersection cardinality in \cite{DGT12, DD15}.

However, such attacks cannot be carried out in the same way using Suggestion 2 if $t$ is at least one. In this case, the citizen will always receive $0$ from the server at the end if she is only inputting a single token. Thus, a corrupt party needs to collect more than $t$ different tokens from a specific person in order to carry out such an attack and learn if another person is reported infected. Furthermore, the threshold $t$ needs not to be publicly known (in fact it may vary depending on the recommendations by the health authorities) but only known by the server making it more difficult for the citizen to know exactly how many tokens she needs to collect in order to perform such an attack.

We remark that there can be privacy concerns about learning the cardinality of the intersection, meaning that there can be both advantages and disadvantages of using Suggestion 1 or 2 corresponding to that the citizen or the server learns $|R(i)\cap L|$. 


\section{Determination of the location of infections}\label{sec:4}

The task is to investigate if there are locations in the region $A$ where it is more likely that citizens get infected. For instance, does infections occur more frequently in the local supermarket than in the church?

We propose to combine the approach in section \ref{sec:2} with the approach in section \ref{sec:3} to determine the location of infections. To this end, we consider the partition of $A$ introduced in section \ref{sec:2} and aim to count the number of infections in each location $A_j$. Since we have no demands for the partitioning other than the ones in \eqref{eq:parts}, the locations could be designed such that the location of interest (e.g. supermarkets, churches, schools, large workplaces, sport facilities etc.) fills up one location. In this way, our method helps determining if certain locations are more likely to let the disease spread.

\subsection{Problem Formulation}
Let $Q \subset I$ be an index set - the set of infected citizens. 
Suppose that citizen $i \in I$, in location $A_k$, has a contact with citizen $j \in I$ who later is tested positive for the disease. If $i$ afterwards is also tested positive, we assume that she was infected by $j$ in location $A_k$. Potentially, $i$ can have been in contact with other infected citizens in locations $A_{k'}$, in which case we choose to view each distinct $A_{k'}$ as a potential infection site and therefore count in each of them. In continuation, we define the set $\A_i$ consisting of the indices $k$ of locations $A_k$, where $i$ has been in contact with an infected citizen. To this end, we denote by $v(\A_i) \in \mathbb{Z}^{M}$ the vector with entries 
\begin{equation}
    v_h(\A_i) = \begin{cases} 1 & \text{if } h \in \A_i \\ 0 & \text{otherwise,} \end{cases}
\end{equation}
where $v_h(\A_i)$ is the $h$'th entry of the vector $v(\A_i)$.
The problem is then to calculate
\begin{equation} \label{eq:sum4}
  Y =  \sum_{i \in Q} v(\A_i). 
\end{equation}


\subsection{Protocol}
We propose to combine the solution to \textbf{(P1)} and \textbf{(P2)}. That is, we assume that each citizen $i$ stores the sets $L(i)$ and $R(i)$, which are introduced in section \ref{sec:3}. Furthermore, we introduce a third set $RA(i)$ that keeps track of which location $A_k$ citizen $i$ was located in when receiving the token $r \in R(i)$. Specifically, $RA(i)$ is the set consisting of the pairs $(r, A_k)$, for all tokens $r$ received by $i$. To this end, we propose that if citizen $i$ is tested positive for the disease, both $L(i)$ and $RA(i)$ is uploaded to the server. This will reveal the locations of where citizen $i$ has been in contact with other citizens, which does leak some private information about $i$. However, by designing the locations $A_k$ such that locations consisting of residences are large and places of interest are small, we believe that it will not be possible to infer the identity of $i$ from $L(i)$ and $RA(i)$. 

The server can then check whether any tokens $r$ received by $i$ in location $A_k$ is in $L$, which would imply that $i$ was infected by the citizen transmitting $r \in L$ in location $A_k$. If multiple tokens $r' \in RA(i)$ is in $L$ and all is received in distinct locations, we do not know where $i$ was infected, but each of the corresponding locations is a potential infection site. Hence, we propose to count all locations as a possible infection site. The proposed method is formally presented in \proref{pro3}.

\begin{pro}{ht}
	\caption{Location of Infections -- Suggestion 1}
	\label{pro3}
		\begin{algorithmic}[1]
		    \Statex \textbf{\textit{Each citizen $i \in I$ does:}}
		    \State For each received token, $i$ attach the location $A_k$ of where the token was received, and creates the set $RA(i)$. 
		    \State If citizen $i$ is tested positive for the disease, both $L(i)$ and $RA(i)$ is uploaded to the server.
		    
		    \vspace{0.2cm}
            \Statex \textbf{\textit{The server does:}}
            
            \State
             For all $i \in Q$, compute the set $$\A_i=\{k\mid \exists (r,A_k)\in RA(i)  \text{ and } r\in L\}$$
            \State Compute 
            \begin{equation}
                Y = \sum_{i \in Q} v(\A_i). 
            \end{equation}
            
		\end{algorithmic}
\end{pro}

The correctness of \proref{pro3} is straight forward, as the server directly computes the desired sum from \eqref{eq:sum4}. To see that the protocol preserves the privacy of citizen $i$, consider $L(i)$ and $RA(i)$ received by the server from citizen $i \in Q$. The former is a set of uniformly random numbers, thus these cannot give the server any direct information. However, if the server also receives $L(j)$ and $RA(j)$ from a citizen $j \in Q$ and a token $r$ in $(r,A_k) \in RA(j)$ is also in $L(i)$, then the server knows that $i$ and $j$ has been in contact in location $A_k$. However, this is exactly the information we want the server to learn, to be able to determine the location of infections. Thus, we view it as unavoidable that the server learns this. Moreover, we must keep in mind that the server only learns this information from citizens $i,j \in Q$, i.e., citizens who are infected. 

From $RA(i)$ the server learns the location of where $i$ has made contact with another citizen. Hence, the server learns some of the locations $i$ has been in during the retention time. However, if the locations are designed such that places where only a few people are located are large and places where many people comes and goes are smaller, then the possibility of inferring the identity of citizen $i$ from $RA(i)$ is small.




As an alternative to \proref{pro3} we also present \proref{pro:alternative}. The concept of this protocol is similar. However, when $i$ is reported infected she only sends $RA(i)$ to the server. The server then needs to call a private set intersection functionality, see \funcref{func:PSI_func} for a definition, with citizen $i$ in order to learn $L(i)\cap R$. This is the tokens infected persons have received from $i$, meaning that $i$ has been in contact with an infected citizen when this token was sent. After the server has learned the intersection it can search for the tokens in $RA$ to find the corresponding locations.

In this way \proref{pro:alternative} is maybe more privacy preserving with respect to the citizens since they only needs to send one of the sets. However, it is probably also more communication consuming since the server and the citizen needs to carry out the intersection in a secure way and the server cannot compute it local.

\begin{func}{ht}
    \caption{$\mathcal{F}_{PSI}$}
	\label{func:PSI_func}
		\begin{algorithmic}[1]
		    \Statex On input a set $X$ of cardinality $n$ from the receiver and a set $Y$ of cardinality $m$ from the sender, the functionality outputs $\perp$ to the sender and $X\cap Y$ to the receiver.
		\end{algorithmic}
\end{func}

\begin{pro}{ht}
	\caption{Location of Infections -- Suggestion 2}
	\label{pro:alternative}
		\begin{algorithmic}[1]
		    \Statex \textbf{\textit{Each citizen $i \in I$ does:}}
		    \State  For each received token, $i$ attach the location $A_k$ of where the token was received, and creates the set $RA(i)$.
		    \State If citizen $i$ is tested positive for the disease, it uploads $RA(i)$ to the server. Furthermore, it engage in $\mathcal{F}_{PSI}$ inputting $L(i)$ with the server. 
		    
		    \vspace{0.2cm}
            \Statex \textbf{\textit{The server does:}}
            
            \State The server inputs $R=\bigcup_{i\in Q}R(i)$ to $\mathcal{F}_{PSI}$ and receives $L(i)\cap R$.
            \State For all $i\in Q$, compute the set $$\A_i=\{k\mid \exists (r,A_k)\in RA \text{ and } r\in L(i)\cap R\}$$
            \State Compute 
            \begin{equation}
                Y = \sum_{i \in Q} v(\A_i). 
            \end{equation}
		\end{algorithmic}
\end{pro}

As a remark, we notice that we can interchange the roles of $R(i)$ and $L(i)$ meaning that it does not matter to which of the sets we attach the location. Thus, in \proref{pro:alternative} we could instead have let the citizens produce $LA(i)$ and send these to the server. The server then stores $LA=\bigcup_{i\in Q}LA(i)$ and inputs $L$ to $\mathcal{F}_{PSI}$ which reveal $R(i)\cap L$ to the server. This modification makes it very similar to \proref{pro:Notify2} where the server learnt $|R(i)\cap L|$.

\section{Determination of the time it takes for the disease to spread}\label{sec:5}

The question is to establish the timeline of how the disease moves across infection sites. For instance, if a substantial amount of infections has occurred in a certain city, how much time does it take before there also is a large amount of infections in the neighboring city.  

We will again use the locations $A_k$ of the area $A$, defined in section \ref{sec:2}. Thus, we seek to determine the number of infections in each location $A_k$ at each time step. This data serves as a timeline of how the disease spreads across the country.

\subsection{Problem Formulation}
The problem turns out to be very similar to \textbf{(P3)}, which is solved in section \ref{sec:4}. Hence, we use the notation introduced in this section and additionally introduce the index set $T=\{t_0,t_1,t_2,\ldots\}$ being the set of time steps where it is desired to measure the evolution of the disease. We denote the set of infected citizens between the time step $t_{h-1}\in T$ and time step $t_{h}\in T$ by $Q_{t_h}$. 

The task is then to calculate the number of infections occurring in each location in the time intervals, i.e.
\begin{equation}\label{eq:Ydelta}
    Y_{t_h} = \sum_{i\in Q_{t_h}} v(\A_i),
\end{equation}

\subsection{Protocol}
We propose to use \proref{pro3} or \proref{pro:alternative} to calculate \eqref{eq:Ydelta} with the distinction that the server must keep track of the time it receives $L(i)$ and $RA(i)$ for $i \in Q$. In this way, the server can produce $Y_{t_h}$ for each $h=1,2,\ldots$. We write the protocol in \proref{pro4} where we have used \proref{pro3} as our starting point.

\begin{pro}{ht}
	\caption{Time and Location of Infections}
	\label{pro4}
		\begin{algorithmic}[1]
		    \Statex \textbf{\textit{Each citizen $i \in I$ does:}}
		    \State $i$ does the same steps as she does in \proref{pro3}.
		    
		    \vspace{0.2cm}
            \Statex \textbf{\textit{The server does:}}
            
            \State Upon receiving $L(i)$ and $RA(i)$ between time step $t_{h-1}$ and $t_h$, add $i$ to $Q_{t_h}$.
            \State For all $i \in Q_{t_h}$, compute the set $$\A_i=\{k\mid \exists (r,A_k)\in RA(i)  \text{ and } r\in L\}$$
            \State Compute at each time $t_h \in T$
            \begin{equation}
                Y_{t_h} = \sum_{i \in Q_{t_h}} v(\A_i). 
            \end{equation}
            
		\end{algorithmic}
\end{pro}
That \proref{pro4} is correct and privacy preserving follows since \proref{pro3} is.

\section{Conclusion}\label{sec:6}
The paper presents privacy preserving methods for answering questions related to the behavior of citizens during an epidemic and related to the spread of the disease. Our methods are based on the observation that a large percentage of the population in most countries carries a smart-phone. To this end, the efficiency of our methods relies on the willingness of citizens to install and use these app-based methods.   







\bibliographystyle{model1-num-names}

\bibliography{sample.bib}







\end{document}